\documentclass[%
 reprint,
nofootinbib,
 amsmath,amssymb,
 aps,
]{revtex4-2}

\usepackage{graphicx}
\usepackage{dcolumn}
\usepackage{bm}
\usepackage{hyperref}

\usepackage{natbib}
\usepackage{graphicx}
\usepackage{amsmath}
\usepackage{amssymb}

\usepackage{slashed}

\newcommand{\beq}{\begin{equation}}
\newcommand{\eeq}{\end{equation}}
\newcommand{\ret}{\mathrm{ret}}
\newcommand{\trento}{\texttt{T$_\mathtt{R}$ENTo}}

\newcommand{\SMASH}{\texttt{SMASH}}

\newcommand{\urqmd}{\texttt{UrQMD}}

\begin{document}


\title{Polarized Photons from the Early Stages of Relativistic Heavy-Ion Collisions}

\author{Sigtryggur Hauksson}
 \email{sigtryggur.hauksson@ipht.fr}
\affiliation{%
Institut de Physique Théorique, CEA/Saclay, Université Paris-Saclay, 91191, Gif sur Yvette, France
}%

\author{Charles Gale}
 \email{gale@physics.mcgill.ca}
\affiliation{
Department  of  Physics,  McGill  University,  3600  University  Street,  Montréal,  QC,  Canada  H3A  2T8
}%


\begin{abstract}
The polarization of real photons emitted from early-time heavy-ion collisions is calculated, concentrating on the contribution from bremsstrahlung and quark-antiquark annihilation processes at leading order in the strong coupling. The effect of an initial momentum space anisotropy of the parton distribution is evaluated using a model for the non-equilibrium scattering kernel for momentum broadening. The effect on the photon polarization is reported for different degrees of anisotropy. The real photons emitted early during in-medium interactions will be dominantly polarized along the beam axis. 
\end{abstract}

\maketitle


\section{Introduction}

The theory of the nuclear strong interaction, QCD, features a transition from a phase where the relevant degrees of freedom are quarks and gluons -- at high temperature -- to one where the appropriate basis consists of composite hadrons at a lower temperature. This transition has been predicted by several theoretical approaches, including the non-perturbative field-theoretical framework of lattice QCD \cite{[{See, for example, }][{, and references therein.}]Aarts:2023vsf}. Decades of intense theoretical effort have revealed that the transition from confined hadrons to partons is not a thermodynamic phase transition in the proper sense but rather an analytic crossover \cite{[{See, for example, }][{, and references therein.}]Aoki:2006we}. On the experimental side, this exotic state of strongly interacting matter -- the quark-gluon plasma (QGP) -- has been observed in the relativistic collisions of nuclei (``heavy-ions'') performed at the Relativistic Heavy-Ion Collider (RHIC), and its existence has been later confirmed by experiments performed at the Large Hadron Collider (LHC) \cite{Busza:2018rrf}. There also is strong evidence supporting the presence of QGP  in smaller systems \cite{Strickland:2018exs}. 

Even though new aspects of the QGP are continuously being discovered, it is fair to assert that the field is well poised to enter a phase of quantitative characterization, owing in large part to the large variety of experimental observables being measured by the experimental collaborations. Measurements designed to probe the QGP have reported results on soft hadron collective behavior \cite{Snellings:2006qw}, on QCD jet modification and energy loss \cite{Connors:2017ptx}, on photon \cite{David:2019wpt} and dilepton \cite{Salabura:2008zz} production, and on many other aspects as well. The theoretical tools developed to study the dynamics of nuclear collisions and the formation of the QGP typically consist of multistage models, rendered necessary by the complexity of the nuclear reaction. Recent examples of such a composite theoretical approach are studies \cite{Heffernan:2023utr,JETSCAPE:2020shq,JETSCAPE:2020mzn,Bernhard:2019bmu} where the initial state model (e.g. \trento~\cite{trento}, IP-Glasma \cite{Schenke:2012wb}) preceded a fluid-dynamical phase (\cite{Schenke:2010nt,Schenke:2010rr,Paquet:2015lta,Shen:2014vra}. A hadronic cascade afterburner (e.g. \urqmd~\cite{Bleicher:1999xi}, \SMASH~\cite{smash}), evolve the final state hadrons until their measurement. Such composite models can make statistically significant statements about transport parameters such as shear and bulk viscosity, as well as quantify the energy loss of energetic QCD jets \cite{JETSCAPE:2021ehl}.

As the multistage modeling of relativistic heavy-ion collisions covers a variety of dynamical conditions ranging from far from equilibrium initial states to almost ideal fluid dynamics, it is important to critically examine its different eras.  In searching for observables capable of revealing the different modeling epochs, penetrating probes such as real and virtual photons impose themselves. Electromagnetic variables are emitted at all stages of the collision and as such can report  on local conditions at their creation point \cite{Gale:2009gc}. The largest uncertainty in the chain of models currently lie at the beginning, in the time span preceding ``hydrodynamization''. Early in the history of the collision, photon emission is liable to occur in media far from equilibrium which necessitate a dedicated theoretical treatment. The emission of photons from non-equilibrium environments has received some recent attention \cite{Hauksson:2017udm,Hauksson:2020wsm}. 

This study will consider photon emission in the early stages of heavy-ion collisions, and will focus more specifically on the polarization states of those photons as a probe of the medium at early times. As our calculation only relies on the medium having a pressure anisotropy, or equivalently a momentum anisotropy in parton distribution, it holds both before hydrodynamization as well as in the beginning of the hydrodynamic stage while pressure anisotropy still persists. Some previous estimates for photon polarization at early times considered leading order direct photon production channels  like those of the Compton process and $q \bar{q}$ annihilation \cite{Baym:2014qfa,Schenke:2006yp,Bhattacharya:2015ada}.  It is known that an equally important contribution as those two -- at the same order in $\alpha_s$, the strong coupling constant -- is that associated with the Landau-Pomeranchuk-Migdal effect (LPM) \cite{Arnold:2001ba,Arnold:2001ms,Arnold:2002ja}. That contribution, evaluated for a medium out of equilibrium forms the basis of this work. It is fair to remind readers that the measurement of real photon polarization states is challenging, owing to the complications related to the external conversions into lepton pairs. The angular distribution of this pairs will reflect the polarization state.  A more realistic proposition is a measurement of virtual photon polarization states, as measured through an internal conversion process leading to a dilepton final state. Consequently, the goal of our work is to first set the foundations for subsequent such evaluations and to perform a first estimate of the polarization signature of an early, non-equilibrium, strongly interacting medium.

Our paper is organized as follows: Section \ref{PolPhotons} lays out the building blocks of our non-equilibrium formalism. The section following that one discusses the collision kernel used to model the medium interactions. The numerical methods used to obtain the production rate of polarized photons are discussed in Section \ref{NumMethods}. Results and conclusion constitute Sections \ref{Results} and \ref{Conclusion}, respectively. We present analytical and numerical details in the Appendices. 

\section{Polarized photon emission}
\label{PolPhotons}

\begin{figure}
\includegraphics[scale=0.7]{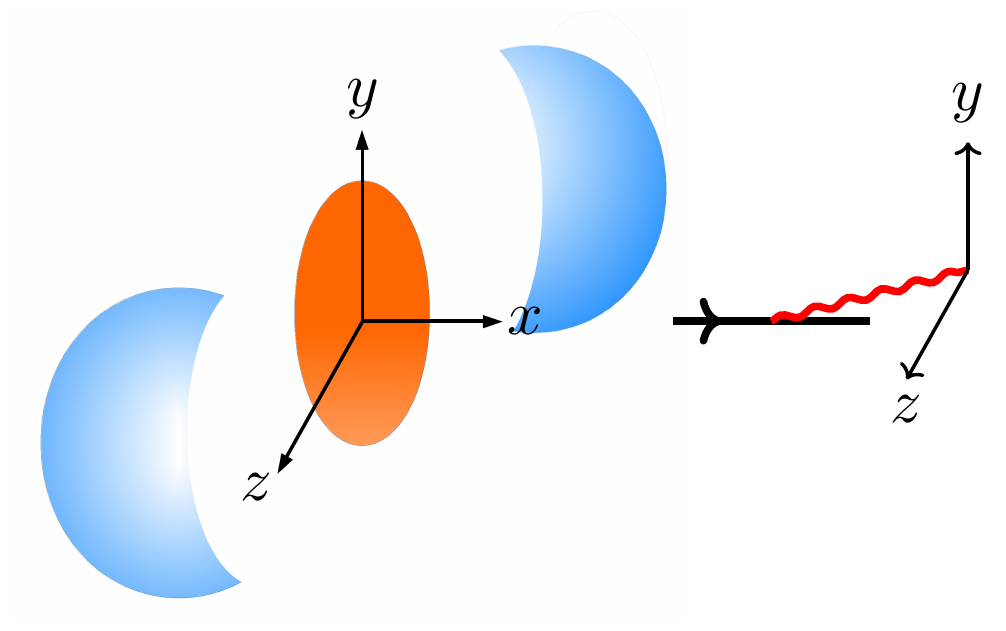}
\caption{Our choice of coordinate system. The \(z\)-axis is chosen to be along the beam axis. For a photon emitted at midrapidity we choose the \(x\)-axis to be along its momentum. The photon (in red) can either be transversely polarized along the beam axis (\(z\)-axis) or transversely polarized orthogonal to the beam axis, i.e. along the \(y\)-axis.}
\label{fig:coordinates}
\end{figure}

\begin{figure}
\includegraphics{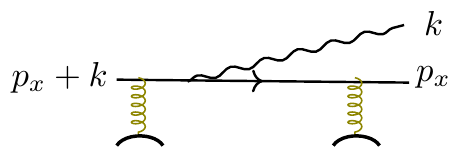}
\caption{Photon radiation through medium-induced bremsstrahlung off a quark}
\label{fig:bremsstrahlung}
\end{figure}

\begin{figure}
\includegraphics{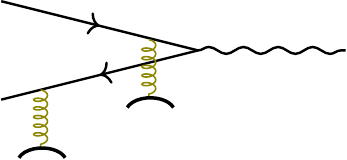}
\caption{Photon radiation through medium-induced quark-antiquark pair annihilation }
\label{fig:pair_ann}

\end{figure}

The quark-gluon plasma radiates photons through two different processes at leading order in perturbation theory. (See \cite{Ghiglieri:2013gia} for higher order corrections.) These processes are two-to-two scattering with a photon in the final state \cite{Baier:1991em,Kapusta:1991qp}, and bremsstrahlung and pair annihilation with a resulting photon \cite{Arnold:2001ba, Arnold:2002ja}. Two-to-two scattering on one hand and bremsstrahlung and pair annihilation on the other hand give roughly equal contribution to photon yield in the plasma \cite{Arnold:2001ms}. While photons from two-to-two scattering have been studied extensively in an anisotropic medium 
and have been shown to be polarized \cite{Baym:2014qfa}, bremsstrahlung and pair annihilation photons have been studied much less in an anisotropic plasma due to the more complicated physics involved. We consider this  now. Polarized photon emission has also been studied in other contexts, including in holography where a background magnetic field is included \cite{Yee:2013qma,Wu:2013qja,Avila:2021rcu}, due to vortical flow in the plasma \cite{Dong:2021fxn}, and due to the chiral magnetic effect \cite{Mamo:2013jda,Ipp:2007ng}. Dilepton polarization has furthermore been considered in \cite{Baym:2017qxy,Shuryak:2012nf,Speranza:2018osi}


 In bremsstrahlung an on-shell photon is emitted collinearly off a quark or an antiquark, see Fig. \ref{fig:bremsstrahlung}. In vacuum this process would be kinematically forbidden as an on-shell quark cannot emit an on-shell photon. However, in a medium the process is made possible to due soft gluon kicks from the medium that bring the quark slightly off-shell. These kicks have momentum \(\sim g \Lambda\) where \(\Lambda\) is a hard scale akin to temperature and \(g\) is the coupling constant. The off-shellness of the quark is therefore of order  \(P^2 \sim g^2 \Lambda^2\) meaning that the emission takes time \(t \sim p/P^2 \sim 1/g^2\Lambda\) where \(p\) is the quark momentum. During this time the quark can receive arbitrarily many soft gluon kicks since the mean-free time between two such kicks is also of order \(1/g^2\Lambda\). All of these kicks need to be included at leading order in perturbation theory \cite{Arnold:2001ba, Arnold:2002ja, Aurenche:1998nw}. In an analogous fashion a quark-antiquark pair can annihilate and radiate a photon due to medium kicks, see  Fig. \ref{fig:pair_ann}.


We will now show that photons emitted through bremsstrahlung and pair annihilation  are polarized in an anisotropic medium. This polarization can be described  by extending the framework developed in  \cite{Arnold:2001ba, Arnold:2002ja, Hauksson:2017udm}, see App. \ref{sec:rate_eq}. To fix ideas we choose the coordinate system in Fig. \ref{fig:coordinates}. The \(z\)-axis lies along the beam axis in a heavy-ion collision. We consider a photon at midrapidity and orient the coordinate system so that the \(x\)-axis lies along its momentum \(\mathbf{k}\). The momentum of the photon can of course have any direction in the plane transverse to the beam axis; aligning it with the short axis of the plasma as in Fig. \ref{fig:coordinates} is simply for illustration. \footnote{In this work, the net polarization of photons emitted from a fluid cell is independent of the angular orientation in the transverse plane as we focus on the effect of longitudinal expansion. This could be generalized to include transverse expansion which breaks this symmetry. Our formalism could also easily be extended to photons at finite rapidity.}  As an on-shell photon is transversely polarized, the polarization basis can be choosen as \(\epsilon_z = (0,0,0,1)\) and \(\epsilon_y = (0,0,1,0)\). The photon is thus polarized along the beam axis or transverse to the beam axis.

In Appendix \ref{sec:rate_eq} we show that the rate of producing \(z\)-polarized photons with momentum \(k\) through bremsstrahlung is
\begin{align}
\label{Eq:rate_z}
&k \frac{d\Gamma_z}{d^3\mathbf{k}} = \frac{6 \alpha_{EM} \sum_s q_s^2}{(2\pi)^3}\, 2 \int_{0}^{\infty} dp_x\;\frac{k}{8p_x^2(k+p_x)} \nonumber \\
&\times  n_f(k+p_x)    \left[ 1-n_f(p_x)\right] \left( F_{\mathrm{in}}(\zeta) A_z + F_{\mathrm{out}}(\zeta) A_y \right)  
\end{align}
where momenta are defined in Fig. \ref{fig:bremsstrahlung} and \(\zeta = k/(p_x+k)\) is the momentum fraction of the photon. 
Similarly, the rate of producing \(y\)-polarized photons is 
\begin{align}
\label{Eq:rate_y}
 &k\frac{d\Gamma_y}{d^3\mathbf{k}}  = \frac{6 \alpha_{EM} \sum_s q_s^2}{(2\pi)^3} 2 \int_{0}^{\infty} dp_x\;\frac{k}{8p_x^2(k+p_x)} \nonumber\\ 
&\times n_f(k+p_x) \left[ 1-n_f(p_x)\right] \left( F_{\mathrm{in}}(\zeta) A_y + F_{\mathrm{out}}(\zeta) A_z \right) 
\end{align}
where \(A_z\) and \(A_y\) have been interchanged relative to Eq. \eqref{Eq:rate_z}. Here \(A_z\)  and \(A_y\) quantify the amount of momentum broadening in the \(z\)- and \(y-\)direction and are defined as
\begin{eqnarray}
\label{Eq:A_zy}
A_z &&=  \mathrm{Re}\, \int \frac{d^2 \mathbf{p}_{\perp}}{(2\pi)^2} \;2p_{z} f_z(\mathbf{p}_{\perp}) \nonumber \\
A_y &&=  \mathrm{Re}\, \int \frac{d^2 \mathbf{p}_{\perp}}{(2\pi)^2} \;2p_{y} f_y(\mathbf{p}_{\perp})
\end{eqnarray}
 where
 \(\mathbf{f}\) solves an integro-differential equation given below. Furthermore, 
\beq
F_{\mathrm{in}}(\zeta) = \frac{(2-\zeta)^2}{\zeta}
\eeq
and
\beq
F_{\mathrm{out}}(\zeta) = \zeta,
\eeq
are polarized splitting functions \cite{Ellis:1996mzs}.
Finally, there are momentum factors \(n_f\) for the incoming and outgoing quarks.  \footnote{We have assumed that there is no chiral imbalance in the medium and that the baryon chemical potential vanishes, so that quarks and antiquarks of both helicities have the same momentum distribution \(n_f\).} The analogous expressions for quark-antiquark pair annihilation are given in Appendix \ref{sec:rate_eq}.

\begin{figure}
\includegraphics{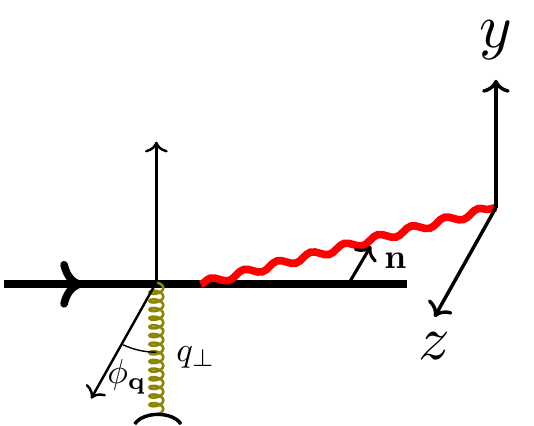}
\caption{Definition of the quantities \(\mathbf{n}\) (the vector between the outgoing quark and photon) and \(\phi_{\mathbf{q}}\) (the angle defining a soft gluon kick of magnitude \(q_{\perp}\)). The photon can be polarized in the \(y\)- or the \(z\)-directions which are transverse to its direction of motion, see also Fig. \ref{fig:coordinates}.}
\label{fig:coordinates_2}

\end{figure}

Eqs. \eqref{Eq:rate_z} and \eqref{Eq:rate_y} for polarized photon emission can be understood intuitively. We consider a \(z\)-polarized photon for concreteness. As the photon travels in the \(x\)-direction, the splitting plane of the photon and the outgoing quark is spanned by \(\widehat{\mathbf{e}}_x\) and a vector orthogonal to \(\widehat{\mathbf{e}}_x\) which we call \(\widehat{\mathbf{n}}\), see Fig. \ref{fig:coordinates_2}.   Eqs. \eqref{Eq:rate_z} and \eqref{Eq:rate_y} show that we can project the vector \(\widehat{\mathbf{n}}\) on to the \(y\)- and the \(z\)-axes and sum  over the contributions.  For \(\widehat{\mathbf{n}}\) in the \(z\)-direction, the \(z\)-polarized photon is polarized in the splitting plane and the hard splitting function is \(F_{\mathrm{in}}\). Momentum broadening is quantified by the \(z\)-component \(A_z\). On the other hand, for \(\widehat{\mathbf{n}}\) in the \(y\)-direction, the \(z\)-polarized photon is polarized out of the splitting plane and the hard splitting function is \(F_{\mathrm{out}}\). Momentum broadening is then  quantified by \(A_y\).

The rate equations for photon emission depend on the function \(\mathbf{f} = (f_z,f_y)\) which quantifies momentum broadening. It solves the integro-differential equation
\beq
\label{Eq:int_eq_full}
2\mathbf{p}_{\perp} = i \delta E \mathbf{f}(\mathbf{p}_{\perp}) + \int \frac{d^2 q_{\perp}}{(2\pi)^2} \;\mathcal{C}(\mathbf{q}_{\perp}) \left[\mathbf{f}(\mathbf{p}_{\perp}) - \mathbf{f}(\mathbf{p}_{\perp} + \mathbf{q}_{\perp}) \right]
\eeq
where
\beq
\delta E = \frac{k}{2p(p+k)} \left[ p_{\perp}^2 + m_{\infty}^2\right].
\eeq
The central ingredient in this equation is the collision kernel \(\mathcal{C}(\mathbf{q}_{\perp})\) which gives the rate for a quark to receive soft gluon kicks of transverse momentum \(\mathbf{q}_{\perp}\). The collision kernel gives rise to a gain term and a loss term in Eq. \eqref{Eq:int_eq_full}.

In an isotropic medium, the collision kernel is by definition isotropic, \(\mathcal{C}(\mathbf{q}_{\perp}) = \mathcal{C}(q_{\perp})\) and one can show that \(\mathbf{f} = \mathbf{p}_{\perp} \widehat{f}(p_{\perp})\). This means that \(A_z = A_y\) and thus there is no net polarization of photons emitted. In an anisotropic medium, \(\mathcal{C}(\mathbf{q}_{\perp})\) depends not only on the magnitude of the kick \(q_{\perp}\) but also on its orientation. 
 In other words, when writing
\beq
\mathbf{q}_{\perp} = \left( q_z, q_y \right) = q_{\perp} (\cos \phi_{\mathbf{q}}, \sin \phi_{\mathbf{q}})
\eeq
the collision kernel  depends on both \(\phi_{\mathbf{q}}\) and \(q_{\perp}\), see Fig. \ref{fig:coordinates_2}. This leads to \(\mathbf{f}\) having more complicated angular dependence so that \(A_z \neq A_y\). Therefore, photon emission from an anisotropic medium is polarized.

\section{Model of the collision kernel in an anisotropic plasma}

As previously argued, the collision kernel for soft gluon kicks \(\mathcal{C}(\mathbf{q}_{\perp})\) is anisotropic in heavy-ion collisions which leads to polarization of photons emitted through bremsstrahlung. The ultimate source of the anisotropy in the kernel is longitudinal expansion of the medium along the beam axis. 
Such a longitudinal expansion gives pressure anisotropy at early and intermediate times with longitudinal pressure \(P_L\) less than transverse pressure \(P_T\)
. On a microscopic level, this means that quark and gluon quasiparticles have an anisotropic momentum distribution with \(\langle p_z^2 \rangle < \langle p^2_x\rangle, \langle p^2_y\rangle\). This is captured by the distribution introduced in Ref. \cite{Romatschke:2003ms}:  
\beq
\label{Eq:RSdistr}
f(\mathbf{p}) = \sqrt{1+\xi} f_{\mathrm{iso}}(\sqrt{\mathbf{p}^2 + \xi p_z^2})
\eeq
where \(f_{\mathrm{iso}}\) is an isotropic distribution, and \(\xi>0\) quantifies the degree of anisotropy. The prefactor \(\sqrt{1+\xi}\) ensures that the number density of quarks and gluons is the same as in equilibrium.  A simple calculation gives the pressure anisotropy \(P_T / P_L\) in terms of the momentum space anisotropy \(\xi\), linking the macroscopic and microscopic descriptions.\footnote{ Specifically, \(P_T/P_L = \frac{1}{2}\frac{\sqrt{\xi} + (\xi-1)\arctan \sqrt{\xi} } { \arctan \sqrt{\xi} - \sqrt{\xi}/(1+\xi) } \), see e.g. \cite{Martinez:2010sc}.}

Ideally, one would want to calculate the collision kernel for momentum broadening, \(\mathcal{C}(\mathbf{q}_{\perp})\), directly from Eq. \eqref{Eq:RSdistr}.
Such a calculation would use that the hard quasiparticles in Eq. \eqref{Eq:RSdistr} radiate soft gluons which are then responsible for momentum broadening. 
This would allow to quantify the degree of photon polarization in a non-equilibrium medium from first principles. Unfortunately, going from Eq. \eqref{Eq:RSdistr} to the collision kernel is difficult in practice, partially due to instabilities that can be present in an anisotropic plasma \cite{Hauksson:2021okc}. 


In this study, we will use a simple model of the collision kernel in a longitudinally expanding medium.   
We take inspiration from results for the collision kernel 
in thermal equilibrium at leading order in perturbation theory \cite{Aurenche:2002pd},
\beq
\label{Eq:eq_kernel}
 \mathcal{C}_{\mathrm{eq}}(\mathbf{q}_{\perp}) = g^2 C_F T \left(\frac{1}{q_{\perp}^2} - \frac{1}{q_{\perp}^2 + m_{D0}^2} \right) 
 \eeq
 Here \(m_{D0}^2\) is the equilibrium Debye mass which describes screening of electric fields. (The equilibrium kernel has furthermore been evaluated at next-to-leading order \cite{Caron-Huot:2008zna}, as well as on the lattice, see e.g. \cite{Panero:2013pla,Moore:2021jwe}.)
In our anisotropic model, we replace 
the equilibrium Debye mass by its anisotropic extension\footnote{In \cite{Romatschke:2003ms} this quantity is referred to as \(m_+^2\).  We have expanded in small \(\xi\) in Eq. \eqref{Eq:Cb_model} but this is simply for convenience and not fundamental to the setup we use.} found in \cite{Romatschke:2003ms}
\beq
\label{Eq:Debye_aniso}
m_D^2(\phi_{\mathbf{q}}) = \left( 1 - \frac{2\xi}{3}\right) m_{D0}^2 + \xi m_{D0}^2 \cos^2 \phi_{\mathbf{q}}.
\eeq
so that 
\beq
\label{Eq:Cb_model}
 \mathcal{C}(\mathbf{q}_{\perp}) = g^2 C_F \Lambda \Big(\frac{1}{q_{\perp}^2} -\frac{1}{q_{\perp}^2+m_D^2(\phi_{\mathbf{q}})} \Big) 
\eeq
The anisotropic correction has an angular dependence with more broadening in the \(z\)-direction than in the \(y\)-direction.  To simplify calculations, we expand the collision kernel in \(\xi\), writing
\begin{align}
\label{Eq:Cb_model}
 \mathcal{C}(\mathbf{q}_{\perp}) 
 \approx g^2 C_F \Lambda &\left(\frac{1}{q_{\perp}^2} - \frac{1}{q_{\perp}^2 + m_{D0}^2}\right.\nonumber\\
 &+ \left. \frac{-2\xi m_{D0}^2/3 + \xi m_{D0}^2 \cos^2 \phi_{\mathbf{q}}}{\left(q_{\perp}^2 + m_{D0}^2\right)^2}\right)
\end{align}
where \(\Lambda\) is  a hard scale, akin to temperature.

Eq. \eqref{Eq:Cb_model} is a toy model for the collision kernel, intended to illustrate how an anisotropic kernel leads to polarized photon emission and to estimate the magnitude of this effect. This toy model only includes changes to the screening of electric fields in an anisotropic medium 
and not the myriad other non-equilibrium effects that can arise.  Nevertheless, this collision kernel can be motivated by theoretical arguments making us belief that it captures some of the salient features of the full non-equilibrium kernel.

In general the collision kernel is defined by
\beq
\label{Eq:coll_def}
\mathcal{C}(\mathbf{q}_{\perp}) = g^2 C_F \int \frac{dq^0 dq_{x} }{(2\pi)^2}\; D_{rr}^{\mu\nu}(Q) v_{\mu} v_{\nu} \,2\pi\delta(v\cdot Q)
\eeq 
where \(v^{\mu} = (1,1,0,0)\) is the four-velocity of the quark emitting a photon. The kernel depends on the statistical correlator for gluons in the medium, 
\beq
\label{Eq:Drr_def}
D_{rr}(Q) := \frac{1}{2} \left\langle\left\{A, A\right\}\right\rangle(Q) = D_{\mathrm{ret}}(Q) \Pi(Q) D_{\mathrm{adv}}(Q)
\eeq
which characterizes the occupation density of a pair of soft gluons. We have omitted Lorentz indices for simplicity. This statistical correlator contains information on how  the soft gluons are emitted by hard quasiparticles with rate \(\Pi(Q)\) and then propagate in the medium according to the retarded propagator \(D_{\mathrm{ret}}(Q) = i/(Q^2-\Pi_{\mathrm{ret}})\) and the advanced propagator \(D_{\mathrm{adv}} = - D_{\mathrm{ret}}^*\).

Making some  heuristic assumptions allows one to employ a sum rule in \cite{Aurenche:2002pd} to motivate the toy model for the collision kernel in Eq. \eqref{Eq:Cb_model}, starting from the definitions in Eqs. \eqref{Eq:coll_def} and \eqref{Eq:Drr_def}. The goal is to include anisotropic corrections to the screening of chromoelectric fields, while ignoring anisotropic corrections to the density of gluons and to change in polarization that occurs during propagation. We work strictly at small anisotropy \(\xi \ll 1\), including only effects of order \(\mathcal{O}(\xi)\).

The first heuristic assumption is to employ the identity \(\Pi(Q) = \frac{\Lambda}{q^0} 2\mathrm{Im}\,\Pi_{\mathrm{\ret}}\)  which is known as the KMS identity and which expresses detailed balance between production and decay of soft gluons. This is not strictly valid in a non-equilibrium medium and amounts to ignoring anisotropic corrections to the density of gluons.
Then one can write
\beq
\label{Eq:rr_KMS}
D_{rr}(Q) :=  \frac{\Lambda}{q^0} \left( D_{\mathrm{ret}} - D_{\mathrm{adv}}\right).
\eeq
At small anisotropy the retarded propagator in Eq. \eqref{Eq:rr_KMS} is
\beq
D_{\ret}^{\mu\nu}(Q) \approx \frac{P_T^{\mu\nu}}{Q^2 - \Pi_T} + \frac{P_L^{\mu\nu}}{Q^2 - \Pi_L}.
\eeq
Here we have only included anisotropic corrections to the screening as given by \(\Pi_T\) and \(\Pi_L\),  see App. \ref{sec:app_sum_rule}.

Our second heuristic approximation is to focus on the anisotropic correction to \(\Pi_L\) and ignore those in \(\Pi_T\). Comparing with the equilibrium calculation \cite{Aurenche:2002pd}, this amounts to calculating anisotropic corrections to the term \(1/(q_{\perp}^2 + m_{D0}^2)\) in Eq. \eqref{Eq:eq_kernel} while leaving the term \(1/q_{\perp}^2\) as is.  This means that we include anisotropic corrections to the screening of chromoelectric fields as given by a Debye mass but do not include anistropic corrections to the screening of chromomagnetic fields.

The reason we use this approximation is that the sum rule we employ does not work for the transverse screening in \(\Pi_T\). This is because the term \(1/(Q^2 - \Pi_T)\) has a pole in the upper half complex plane of \(q_0\) corresponding to Weibel instabilities \cite{Mrowczynski:2016etf}. A formal use of the sum rule would lead to a contribution of the form \(1/(q_{\perp}^2 - m^2)\) which is ill-defined at \(q_{\perp} = m\). The solution to this issue is to use a retarded propagator for soft gluons that includes the mechanism by which non-Abelian interaction caps the growth of the unstable soft gluon modes. This is beyond the scope of this project.


Given these two approximations, one can use the sum rule in \cite{Aurenche:2002pd} nearly directly. The longitudinal retarded self-energy at small anisotropy \(\xi\) is
\begin{align}
\label{Eq:PiL_aniso}
\Pi_L(Q) = \Pi_L^0(x) + \xi \left[\frac{1}{6} \left(1 + 3 \cos 2\theta_n \right) \frac{Q^2}{\mathbf{q}^2} m_{D0}^2 \right. \nonumber \\
\left. +  \Pi_L^0(x)\left( \cos 2\theta_n  - \frac{x^2}{2}\left( 1+3\cos 2\theta_n\right) \right)\right] 
\end{align}
where \(\theta_n\) is the angle between \(\mathbf{q}\) and the anisotropy vector \(\mathbf{n} = \mathbf{e}_z\) which defines a preferred direction in Eq. \eqref{Eq:RSdistr} \cite{Romatschke:2003ms}. Here 
\beq
\Pi_L^0(x) = \left(x^2 -1 \right) m_{D0}^2 \left[\frac{x}{2} \log \frac{x+1 + i\epsilon}{x - 1 + i\epsilon} - 1\right]
\eeq
is the equilibrium value. We next do a  change of variables in Eq. \eqref{Eq:coll_def} to \(x = q_0/q = q_x/\sqrt{q_x^2 + q_{\perp}^2}\) so that \(q_x^2 = x^2 q_{\perp}^2/(1-x^2)\) and \(q^2 = q_{\perp}^2/ (1-x^2)\). Then  one should substitute \(\cos 2 \theta_n = 2(q_z/q)^2 -1 =  2 (1-x^2) \cos^2 \phi_{\mathbf{q}} -1\) to get dependence only on \(x\), \(q_{\perp}\)  and \(\phi_{\mathbf{q}}\).  This gives that the longitudinal contribution to the collision kernel is
\beq
\label{Eq:coll_long}
\frac{\Lambda g^2 C_F}{\pi} \int_0^1 \frac{dx}{x}  \frac{\mathrm{Im}\, \widetilde{\Pi}_L(x,\phi_{\mathbf{q}})}{\left(\mathbf{q}_{\perp}^2 + \mathrm{Re}\,\widetilde{\Pi}_L(x,\phi_{\mathbf{q}}) \right)^2 + \left( \mathrm{Im}\,\widetilde{\Pi}_L(x,\phi_{\mathbf{q}})\right)^2}.
\eeq
where
\begin{align}
\label{Eq:PiL_aniso}
&\widetilde{\Pi}_L(x,\phi_{\mathbf{q}}) = \Pi_L^0(x) + \xi \left[\frac{1}{6} \left(1 + 3 \cos 2\theta_n \right) (x^2-1)m_{D0}^2 \right. \nonumber \\
&\left. +  \Pi_L^0(x)\left( \cos 2\theta_n  - \frac{x^2}{2}\left( 1+3\cos 2\theta_n\right) \right)\right] \bigg \rvert_{\cos 2\theta_n =  2 (1-x^2) \cos^2 \phi_{\mathbf{q}} -1}
\end{align}
has no explicit dependence on \(q_{\perp}\).

@article{Aurenche:2002pd,
@article{Aurenche:2002pd,
 An argument nearly identical\footnote{ The retarded propagator \(1/\left[x^2 q_{\perp}^2 - q_{\perp}^2 - (1-x^2)\Pi_L(x,\phi)\right]\) in Eq. 7 in \cite{Aurenche:2002pd} has an extra pole in the upper half complex plane in the anisotropic case. This pole can be seen by taking the limit \(x= k_0/k \rightarrow \infty\) in which case the propagator becomes \(\sim 1/(1-x^2)/(q_{\perp}^2 + \left( \frac{1}{3} - \frac{1}{3} \xi \cos^2 \phi\right) m_{D0}^2 + \frac{\xi}{3} x^2 \cos ^2 \phi \,m_{D0}^2)\) which has a pole which is parametrically of the form \(x \sim \pm i q_{\perp}/m_D \sqrt{\xi}\) and thus far from the real axis when \(\xi \ll 1\). This is not in contradiction with the usual properties of the retarded propagator as we have imposed \(q_0 = q_x\) and then search for poles in \(q_0\). One can then show that the correction due to this pole to the sum rule in Eq. 9 in \cite{Aurenche:2002pd} is \(\mathcal{O}(\xi^{3/2})\) which is subleading to the \(\mathcal{O}(\xi)\) contributions we consider. } 
 to the one in \cite{Aurenche:2002pd} then shows that Eq. \eqref{Eq:coll_long} is 
\begin{align}
&g^2 C_F \Lambda \left[ \frac{1}{q_{\perp}^2 + \lim_{x\rightarrow \infty} \widetilde{\Pi}_L(x,\phi)} - \frac{1}{q_{\perp}^2 + \widetilde{\Pi}_L(0,\phi)}\right]  \nonumber \\
&= - g^2 C_F \Lambda  \frac{1}{q_{\perp}^2 + m_D^2(\phi_{\mathbf{q}})}
\end{align}
since  \( \widetilde{\Pi}_L(0,\phi) = m_D^2(\phi_{\mathbf{q}})\). Our conclusion is therefore that given our heuristic approximations, the collision kernel is given by Eq. \eqref{Eq:Cb_model}. We emphasize that this collision kernel is not intended to capture all of the non-equilibrium physics but to focus on anisotropic corrections to the screening of chromoelectric fields. 


\section{Numerical method}
\label{NumMethods}

We wish to evaluate the rate of polarized photon emission in an anisotropic medium such as that found in a longitudinally expanding quark-gluon plasma. The starting point is Eqs. \eqref{Eq:rate_z} and \eqref{Eq:rate_y} which require solving the integro-differential equation in Eq. \eqref{Eq:int_eq_full} numerically, assuming the model for the collision kernel given in Eq. \eqref{Eq:Cb_model}.
To solve Eq. \eqref{Eq:int_eq_full} we go to impact parameter space, i.e. the space Fourier conjugate to \(\mathbf{p}_{\perp}\). Defining 
\beq
\label{Eq:f_Fourier}
\mathbf{f}(\mathbf{b}) = \int \frac{d^2 p_{\perp}}{(2\pi)^2} \; e^{i \mathbf{p}_{\perp}\cdot \mathbf{b}}\, \mathbf{f}(\mathbf{p}_{\perp}).
\eeq
the equation we wish to solve becomes
\beq
\label{Eq:int_eq_full_transf}
-2i \nabla_b \delta^{(2)}(\mathbf{b}) = \frac{i k}{2p(p+k)} \left[ -\nabla_b^2 + m_{\infty}^2\right] \mathbf{f}(\mathbf{b}) + \mathcal{C}(\mathbf{b}) \mathbf{f}(\mathbf{b}).
\eeq
where 
\beq
\mathcal{C}(\mathbf{b}) = \int \frac{d^2 p_{\perp}}{(2\pi)^2} \; \left[1 -  e^{i \mathbf{p}_{\perp}\cdot \mathbf{b}}\, \right] \mathcal{C}(\mathbf{p}_{\perp}).
\eeq
A straightforward calculation shows that the collision kernel from Eq. \eqref{Eq:Cb_model} 
is
\beq
\label{Eq:Cfull_b}
\mathcal{C}(\mathbf{b}) = \mathcal{C}_0(b) +  \xi \,\mathcal{C}_1^{(a)}(b) + \xi \cos 2\beta\, \mathcal{C}_1^{(b)}(b)
\eeq
in impact parameter space where \(\mathbf{b}= (b_z,b_y) = \left(\cos \beta , \sin \beta \right) b\). 
The terms of the collision kernel are given by
\beq
\mathcal{C}_0(\mathbf{b}) = \frac{g^2 C_F T}{2\pi} \left[ K_0(m_{D0} b) + \gamma_E + \log \frac{m_{D0}b}{2}\right],
\eeq
\beq
\mathcal{C}_1^{(a)}(b)  = g^2 C_F T \frac{1}{8\pi} \frac{ M^2}{m_{D0}^2} \left( m_{D0} b K_{1}(m_{D0} b)  - 1\right)
\eeq
and
\begin{eqnarray}
&&\mathcal{C}_1^{(b)}(b) = g^2 C_F T \frac{M^2 b^2}{4\pi} \Bigg[ \frac{2}{(m_{D0} b)^4}  \nonumber \\
&&- \frac{1}{2 m_{D0} b} K_1(m_{D0} b) - \frac{1}{\left(m_{D0}b\right)^2} K_2(m_{D0} b)\Bigg]
\end{eqnarray}

In order to solve Eq. \eqref{Eq:int_eq_full_transf} we do an expansion in small \(\xi\), giving
\beq
\label{Eq:f_expansion}
\mathbf{f}(\mathbf{b}) = \mathbf{f}_0(\mathbf{b}) + \xi \mathbf{f}_1(\mathbf{b}) + \dots
\eeq
The zeroth order solution  in \(\xi\) satisfies the usual isotropic equation 
\beq
\label{Eq:f0_eq}
 \frac{ik}{2p(p+k)} \left[ -\nabla_b^2 + m_{\infty}^2\right] \mathbf{f}_0(\mathbf{b}) +  \mathcal{C}_0(b) \mathbf{f}_0(\mathbf{b}) = -2i \nabla_b \delta^{(2)}(\mathbf{b})
\eeq
and can be shown to have angular dependence  \(\mathbf{f}_0(\mathbf{b}) \sim (\cos \beta, \sin \beta)\). The first order satisfies
\begin{eqnarray}
\label{Eq:f1_eq}
\frac{ik}{2p(p+k)} && \left[ -\nabla_b^2 + m_{\infty}^2\right] \mathbf{f}_1(\mathbf{b}) +\mathcal{C}_0(b) \mathbf{f}_1(\mathbf{b}) \nonumber \\
&&=   -\left[ \mathcal{C}_1^{(a)}(b) +  \cos 2\beta\, \mathcal{C}_1^{(b)}(b)\right] \mathbf{f}_0(\mathbf{b}).
\end{eqnarray}
Due to the angular dependence of the right hand side we can write in full generality
\beq
f_{1z}(\mathbf{b}) = \cos \beta f_1^{(1z)}(b) + \cos 3 \beta\,f_1^{(3)}(b) 
\eeq
and
\beq
f_{1y}(\mathbf{b}) = \sin \beta f_1^{(1y)}(b) + \sin 3 \beta\,f_1^{(3)}(b) 
\eeq
where these functions solve
\beq
\label{Eq:f11z}
\mathcal{K} \left[ f_1^{(1z)}(\mathbf{b}) \right] 
+\mathcal{C}_0(b) f_1^{(1z)}(\mathbf{b}) =   -\left[ \mathcal{C}_1^{(a)}(b) +  \frac{1}{2} \mathcal{C}_1^{(b)}(b)\right] f_0,
\eeq
\beq
\label{Eq:f11y}
\mathcal{K} \left[ f_1^{(1y)}(\mathbf{b}) \right] 
+\mathcal{C}_0(b) f_1^{(1y)}(\mathbf{b}) =   -\left[ \mathcal{C}_1^{(a)}(b) -  \frac{1}{2} \mathcal{C}_1^{(b)}(b)\right] f_0
\eeq
with
\beq
\mathcal{K} \left[ \mathbf{f}(\mathbf{b})\right] = -\frac{ik}{2p(p+k)} \left[ \frac{d^2}{db^2} + \frac{1}{b} \frac{d}{db} - \frac{1}{b^2} - m_{\infty}^2\right] \mathbf{f}(\mathbf{b}).
\eeq
(The differential equation for \(f_1^{(3)}\) is given in App. \ref{sec:app_num}.)

Our goal is to evaluate 
\beq
\label{Eq:Az_impact}
A_z =  2\mathrm{Im}\, \partial_{b_z} f_{z}(\mathbf{b}) \bigg\rvert_{b=0}  = 2 \mathrm{Im}\,\frac{\mathbf{\widehat{b}}\cdot \mathbf{f}_0 + \xi \,f_1^{(1z)}}{b} \bigg\rvert_{b=0}
\eeq
and 
\beq
\label{Eq:Ay_impact}
A_y =  2\mathrm{Im}\, \partial_{b_y} f_{y}(\mathbf{b}) \bigg\rvert_{b=0}  = 2 \mathrm{Im}\,\frac{\mathbf{\widehat{b}}\cdot \mathbf{f}_0 + \xi \,f_1^{(1y)}}{b} \bigg\rvert_{b=0}
\eeq
 which were defined in Eqs. \eqref{Eq:rate_z} and \eqref{Eq:rate_y}. Thus, we only need to know \(f_1^{(1z)}(b)/b\) and \(f_1^{(1y)}(b)/b\) in the limit \(b\rightarrow 0\) where it must be finite. This gives the boundary condition that  This is done by demanding that \(f_1^{(1z)}\) and \(f_1^{(1y)}\) vanish at \(b=0\). The other boundary condition is that the functions vanish as \(b\rightarrow\infty\) as can be seen from Eq. \eqref{Eq:f_Fourier}.
 
 To evaluate \(f_1^{(1z)}(b)/b\) and \(f_1^{(1y)}(b)/b\) at \(b=0\), we demand that the functions \(f_1^{(1z)}\) and \(f_1^{(1y)}\) vanish at very large \(b\) and evolve the functions numerically to small \(b\) using Eqs. \eqref{Eq:f11z} and \eqref{Eq:f11y}. In practice, this means that we start the evolution at a large but finite value of \(b\) where \(f_1^{(1z)}\) and \(f_1^{(1y)}\) are initialized to a small value. 
A typical numerical solution for \(f_1^{1z}\) and \(f_1^{1y}\) then blows up as evolved towards \(b\rightarrow 0\). We must then extract the finite part of our numerical solution. This is done  by matching with known, analytic solutions of the differential equations in the small \(b\) limit. 

For instance, focusing on Eq. \eqref{Eq:f11z},  we call the particular solution \(w(b)\) and the two independent solutions of the homogeneous equation \(w_1(b)\) and \(w_2(b)\). These are known analytically at small \(b\), see Appendix  \ref{sec:app_num}. We can write our numerical solution in full generality at small \(b\) as 
\beq
f_1^{(1z)}(b) = w(b) + \alpha_1 w_1(b) + \alpha_2 w_2(b)
\eeq
where \(\alpha_1\) and \(\alpha_2\) are found numerically. To extract from this a solution with the right behaviour as \(b \rightarrow 0\), one must in essence subtract a linear combination of \(w_1\) and \(w_2\) which satisfies the boundary condition at \(b\rightarrow \infty\). Then one is left with a solution which satisfies boundary conditions both at \(b=0\) and \(b\rightarrow \infty\) and which gives \(f_1^{(1z)}(b)/b\) at \(b=0\). This procedure is explained in further detail in Appendix \ref{sec:app_num}, see also \cite{Aurenche:2002wq, Aurenche:2002pd,Jeon:2003gi} for earlier work in the isotropic case. A major difference with the isotropic case is that Eqs. \eqref{Eq:f11z} and \eqref{Eq:f11y} have a non-trivial right hand side which complicates the matching procedure. For instance, one must find an analytic solution \(w(b)\) of the full differential equations at small \(b\), including the right hand side. Furthermore, cancellation errors between solutions of the full differential equation and the homogeneous equation must carefully be avoided to get reliable results, see  Appendix \ref{sec:app_num}.


\section{Results}
\label{Results}

\begin{figure}
\includegraphics[scale=0.42]{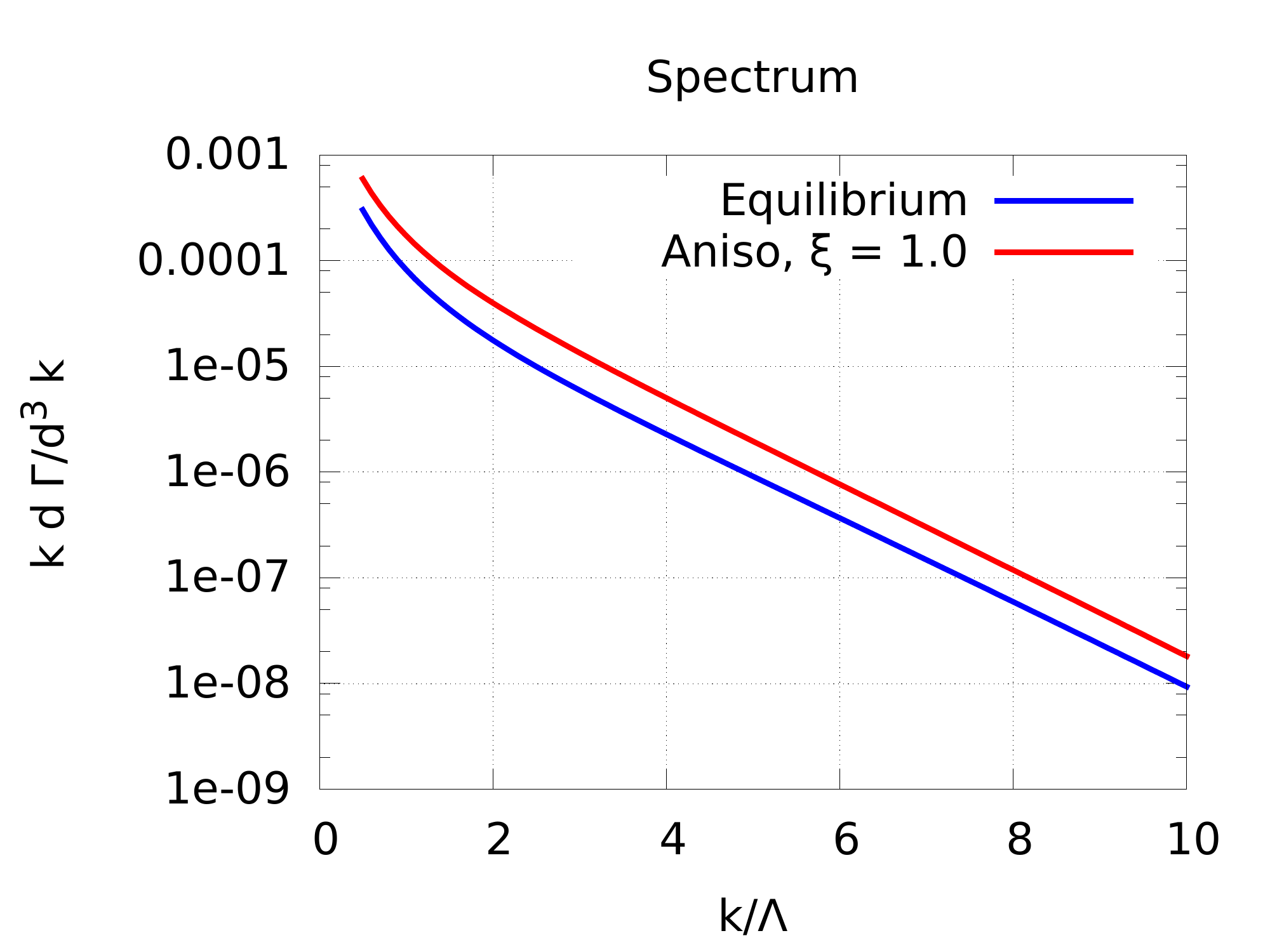}
\caption{Spectrum of photons coming from bremsstrahlung and pair annihilation in a plasma at effective temperature \(\Lambda\). The anisotropic plasma has \(\xi = 1.0\).}
\label{fig:spectrum}
\end{figure}

\begin{figure}
\includegraphics[scale=0.42]{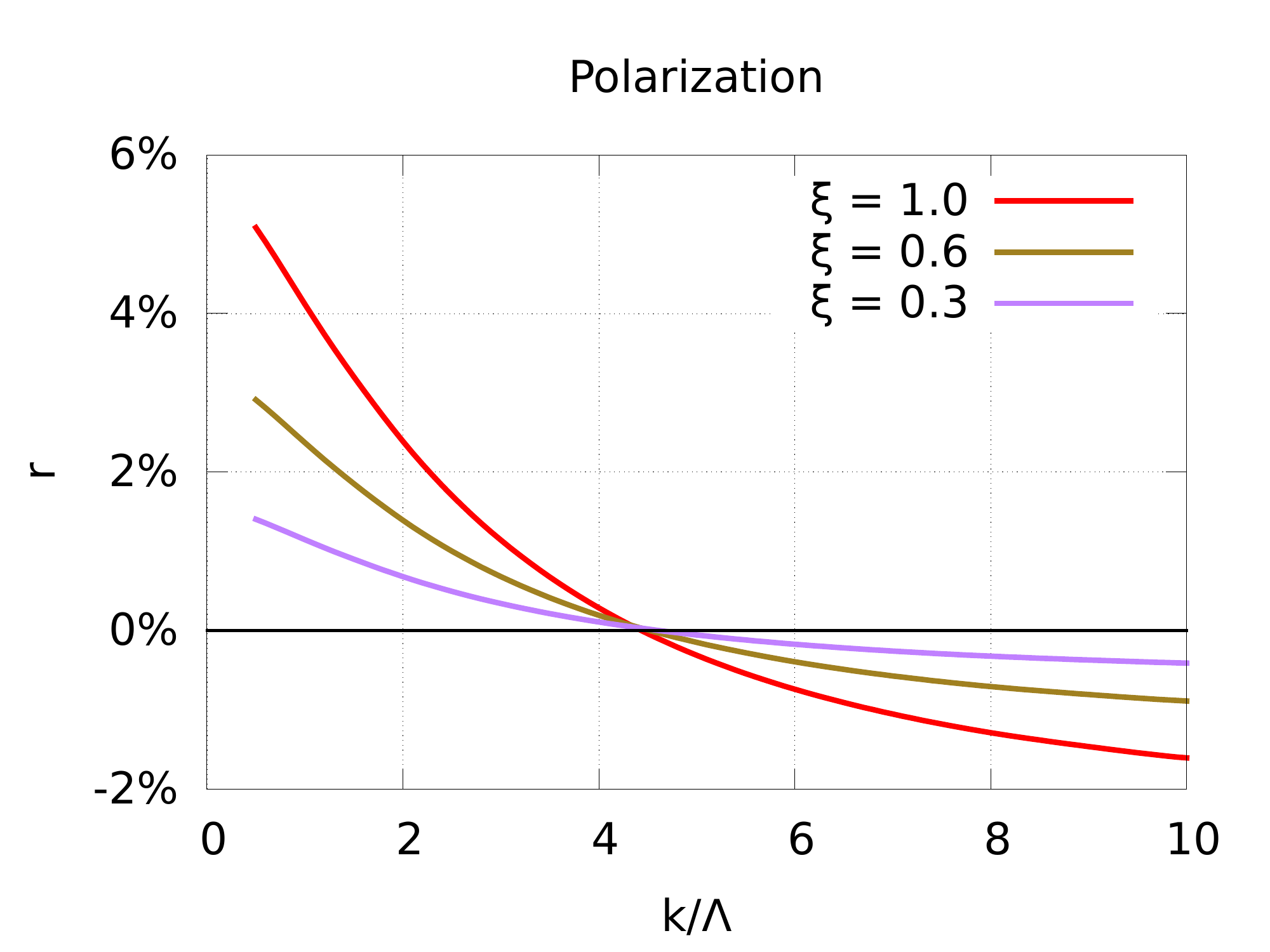}
\caption{Degree of polarization of photons emitted from bremsstrahlung and pair annihilation in an anisotropic plasma with  \(\xi = 1.0\). The quantity \(R\) is defined in Eq. \eqref{Eq:R_def}.}
\label{fig:polarization}
\end{figure}

Figs. \ref{fig:spectrum} and \ref{fig:polarization} are the main results of this work. They show the rate of photon production through bremsstrahlung and pair-annihilation in an anisotropic quark-gluon plasma with fixed anisotropy \(\xi\). The collision kernel is given by Eq. \eqref{Eq:Cb_model} while the momentum distribution of medium quarks is 
\beq
f(\mathbf{p}) =  \frac{\sqrt{1+\xi}}{e^{\sqrt{p^2 + \xi p_z^2}/\Lambda}+1}
\eeq 
where \(\Lambda\) can be seen as an effective temperature.
Fig. \ref{fig:spectrum} shows the total rate for producing photons at mid-rapidity and with momentum \(k\), i.e.
\beq
k \frac{d \Gamma}{d^3 \mathbf{k}} = k \frac{d \Gamma_z}{d^3 \mathbf{k}} + k \frac{d \Gamma_y}{d^3 \mathbf{k}},
\eeq
where \( k \frac{d \Gamma_z}{d^3 \mathbf{k}}\) is the rate of producing photons polarized along the beam axis and \( k \frac{d \Gamma_y}{d^3 \mathbf{k}}\) is the rate for photons polarized orthogonal to the beam axis and to the photon momentum.
Fig. \ref{fig:polarization}  shows the degree of polarization at different momenta defined as 
\beq
\label{Eq:R_def}
R = \frac{ k \frac{d \Gamma_z}{d^3 \mathbf{k}} -  k \frac{d \Gamma_y}{d^3 \mathbf{k}}}{ k \frac{d \Gamma_z}{d^3 \mathbf{k}} +  k \frac{d \Gamma_y}{d^3 \mathbf{k}}}.
\eeq
 These quantities are shown for three values of anisotropy parameter, \(\xi = 0.3\), \(\xi = 0.6\) and \(\xi = 0.9\), which correspond to pressure anisotropy of \(P_L/P_T \approx 0.81\), \(P_L/P_T \approx 0.68 \)  and \(P_L/P_T \approx 0.57\) respectively. These are rather moderate values of pressure anisotropy which can be found in the early and intermediate stages of heavy-ion collisions.

The spectrum in an anisotropic medium is higher than that in an equilibrium medium at the same effective temperature \(\Lambda\), as can be seen in Fig. \ref{fig:spectrum}. This is due to the factor \(\sqrt{1+\xi}\) in the momentum distribution in Eq. \eqref{Eq:RSdistr} which increases the number of quarks with \(p_z = 0\) which can emit photons at midrapidity. This effect is partially compensated by anisotropic corrections which reduce the collision kernel \(\mathcal{C}(\mathbf{q}_{\perp})\) meaning that a given quark receives less momentum broadening, reducing the rate at which it emits photons collinearly. 

 More interesting is the polarization \(R\) as a function of momentum \(k\). As seen in Fig. \ref{fig:polarization}, the polarization has different signs for lower and higher values of the photon energy \(k\): it is along the beam axis for lower values of \(k\) while it is orthogonal to the beam axis at higher values of \(k\). This owes to the interplay between bremsstrahlung and pair annihilation. As shown in Appendix \ref{sec:rate_eq}, because of the different polarized splitting functions, bremsstrahlung tends to give photons polarized along the \(z\)-axis, while pair annihilation tends to give polarization along the \(x\)-axis, see Fig. \ref{fig:coordinates}. As bremsstrahlung is suppressed at high \(k\) (there are few medium quarks with energy higher than \(k\)) \(x\)-polarization dominates in that regime. On the contrary, pair annihilation is suppressed at low \(k\) since the number of quarks with energy less than \(k/2\) is phase space suppressed. This gives \(z\)-polarized photons in that regime.

Despite the complicated dependence of polarization on photon momentum, polarization along the beam axis dominates.  This is simply because there are many more photons at lower \(k\) and thus their polarization is dominant. Furthermore, in \cite{Baym:2014qfa} it was shown that photons from two-to-two scattering, which are equally important as bremsstrahlung and pair annihilation photons, are also predominantly polarized along the beam axis, with an even greater magnitude of polarization. Thus a definite prediction of our work is that medium photons are polarized along the beam axis.

To make contact with potential experiments on photon polarization more work is needed. Firstly, all photon sources that cannot be subtracted in experiments need to be included, such as prompt photons and photons from the hadronic stage. Secondly, calculation of the rate of photon production in an anisotropic medium  need to be folded with hydrodynamic or kinetic theory simulations of the medium to get a realistic evolution of the medium anisotropy.

\section{Conclusion}
\label{Conclusion}

In this work, we have calculated for the first time the degree of polarization for photons emitted in bremsstrahlung and off-shell pair annihilation processes in a hot medium consisting of quarks and gluons.  Our evaluation includes the LPM regime and is at complete leading order in the strong coupling. The polarization of the real photons originating from bremsstrahlung and annihilation processes depends on the anisotropy of the original parton distribution, and therefore  the polarization can instruct us on the dynamics in an environment that is not accessible to the vast majority of probes and observables measured in relativistic heavy-ion collisions. Specifically, it gives a measure of the pressure anisotropy at early times. 

We trust that the methods and techniques developed and used here will be useful in the evaluations of polarization signatures of real and virtual photons, evaluated with scattering kernels for momentum broadening derived from microscopic theories and using time-evolution models based in QCD. 

\acknowledgments
C. G. acknowledges the support of the Natural Sciences and Engineering Research Council of Canada (NSERC), SAPIN-2020-00048.

\appendix

\section{Derivation of rate equations}
\label{sec:rate_eq}

We wish to show that the total rate for polarized photon production through bremsstrahlung and pair annihilation can be written as
\begin{eqnarray}
\label{Eq:Gammaz_app}
k \frac{d\Gamma_z}{d^3\mathbf{k}} &&= \frac{6 \alpha_{EM} \sum_s q_s^2}{(2\pi)^3} \int_{-\infty}^{\infty} dp^z\;
F(\mathbf{k} + \mathbf{p}) \left[ 1-F(\mathbf{p})\right] \nonumber \\
&&\times \frac{1}{2} \frac{1}{4 (p^x)^2 (p^x+k)^2} \left[
(2p^x +k)^2 A_z + k^2 A_y\right]
\end{eqnarray}
and
\begin{eqnarray}
\label{Eq:Gammay_app}
k \frac{d\Gamma_y}{d^3\mathbf{k}} &&= \frac{6 \alpha_{EM} \sum_s q_s^2}{(2\pi)^3} \int_{-\infty}^{\infty} dp^z\;
F(\mathbf{k} + \mathbf{p}) \left[ 1-F(\mathbf{p})\right] \nonumber \\
&&\times \frac{1}{2} \frac{1}{4 (p^x)^2 (p^x+k)^2} \left[
 k^2 A_z +  (2p^x +k)^2 A_y\right]
\end{eqnarray}
where \(A_z\) and \(A_y\) are defined in Eq. \eqref{Eq:A_zy}. Here the momentum distributions are contained in 
\beq
F(p^x) = \theta(p^x) f(\mathbf{p}) + \theta(-p^x) \left( 1-f(-\mathbf{p})\right).
\eeq
 Looking at the momentum factors in Eqs. \eqref{Eq:Gammaz_app} and \eqref{Eq:Gammay_app}, we see that bremsstrahlung off a quark or an antiquark corresponds to \(p^x > 0\) and \(p^x <-k\). These cases can be rewritten to give Eqs. \eqref{Eq:rate_z} and \eqref{Eq:rate_y}. Furthermore, \(-k < p^x < 0\) corresponds to quark-antiquark pair annihilation. As the quark moves in the same direction as the photon we can set e.g. \(f(\mathbf{p}) = f(p^x)\).

The derivation of polarized photon emission in Eqs. \eqref{Eq:Gammaz_app} and \eqref{Eq:Gammay_app} is similar to that for unpolarized emission found in \cite{Arnold:2001ba, Arnold:2002ja, Hauksson:2017udm}. The polarized rate comes from evaluating the diagram in Fig. \ref{fig:app_diagram}.   The gluon ladders which represent soft kicks from the medium are evaluated in the same way as for unpolarized photon emission. Physically, this is because the soft kicks do not have enough energy to flip the helicity of quarks. The gluon ladders are  summed to all orders to give the integral equation in Eq. \eqref{Eq:int_eq_full}. On the contrary, one must keep track of polarization at the hard emission vertices to evaluate the polarized emission rate.

For a bare quark loop without soft gluon kicks, Fig. \ref{fig:app_diagram2}, the hard emission vertices for a \(z-\)polarized photon take the form 
\beq
H^{zz} := \epsilon^z_{\mu} \epsilon^{z\,*}_{\nu} \mathrm{Tr}\, \left[  \gamma^{\mu} (\slashed{K}  + \slashed{P}) \gamma^{\nu} \slashed{P}\right].
\eeq
Here \(\epsilon^z_{\mu}\) is the photon polarization tensor for \(z\)-polarization. This trace is most easily evaluated by using that e.g. 
\beq
\slashed{P} = \sum_s u^s(\mathbf{p}) \overline{u}^s(\mathbf{p})
\eeq
where the sum is over spin states and
\beq
u^s(\mathbf{p}) = \sqrt{2p} 
\begin{bmatrix}
\frac{1-\mathbf{\sigma}\cdot \widehat{\mathbf{p}}}{2} \xi^s \\
\frac{1+\mathbf{\sigma}\cdot \widehat{\mathbf{p}}}{2} \xi^s
\end{bmatrix}
\eeq
are helicity states of quarks \cite{Peskin:1995ev}. Here \(\xi^s\) form a basis for two-component spin states. One can then show by an explicit calculation that
\begin{eqnarray}
\label{Eq:Hzz}
H^{zz} &&=  \sum_{s,t} \left[  \gamma^{z} u^s(\mathbf{p} + \mathbf{k}) \overline{u}^s(\mathbf{p} + \mathbf{k})  \gamma^{z} u^t(\mathbf{p}) \overline{u}^t(\mathbf{p})\right]  \nonumber \\
&&= \frac{8 p (p+k)}{p^2 (p+k)^2} \left[ (2p+k)^2 p_{\perp\,z}^2 + k^2 p_{\perp\,y}^2\right].
\end{eqnarray}
Similarly, the hard emission vertices for emission of \(y\)-polarized photons are
\beq
\label{Eq:Hyy}
H^{yy} = \frac{8 p (p+k)}{p^2 (p+k)^2} \left[ (2p+k)^2 p_{\perp\,y}^2 + k^2 p_{\perp\,z}^2\right].
\eeq
This is the same as Eq. \eqref{Eq:Hzz}, except that \(p_{\perp\,z}^2\) and \(p_{\perp\,y}^2\) have been interchanged.

To include soft gluon as in Fig. \ref{fig:app_diagram} one simply replaces one of the hard vertices by the dressed vertex \(\mathbf{f}(\mathbf{p}_{\perp})\) which includes gluons rungs and obeys the integral equation in Eq. \eqref{Eq:int_eq_full}. This means that in Eqs. \eqref{Eq:Hzz} and \eqref{Eq:Hyy} one replaces \(p_{\perp\,z}^2 \longrightarrow p_{\perp\,z} f_z\) and \(p_{\perp\,y}^2 \longrightarrow p_{\perp\,y} f_y\). This reproduces Eqs. \eqref{Eq:Gammaz_app} and \eqref{Eq:Gammay_app}.

\begin{figure}
\includegraphics[scale=1.5]{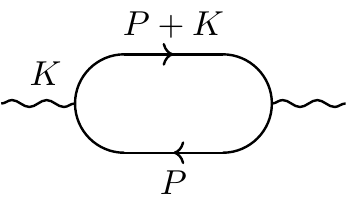}
\caption{Definition of momenta in photon emission.}
\label{fig:app_diagram2}
\end{figure}

\begin{figure}
\includegraphics[scale=1.5]{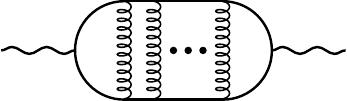}
\caption{Diagram for medium-induced photon emission. The gluon rungs are responsible for momentum broadening.}
\label{fig:app_diagram}
\end{figure}

\section{The retarded gluon propagator at small anisotropy}
\label{sec:app_sum_rule}

In a system whose hard quasiparticles have the momentum distribution in Eq. \eqref{Eq:RSdistr}, the retarded propagator for soft gluons is \cite{Romatschke:2003ms,Hauksson:2021okc}
\begin{align}
D_{\mathrm{ret}}^{\mu\nu}(Q) &= \left(P_T^{\mu\nu} - C^{\mu\nu} \right) D_{\ret}^B \nonumber \\
&+\left[\left( Q^2 - \Pi_c\right) P_L^{\mu\nu} + \left(Q^2 - \Pi_L \right) C^{\mu\nu} + \Pi_d D^{\mu\nu} \right] D_{\ret}^A
\end{align}
where 
\beq
D_{\ret}^A = \frac{1}{\left( Q^2 - \Pi_L\right) \left( Q^2 - \Pi_c\right) - \frac{Q^2\mathbf{q}^2}{q_0^2}\Pi_d^2}
\eeq
and
\beq
D_{\ret}^B = \frac{1}{Q^2 - \Pi_e}.
\eeq
The tensors \(P_T^{\mu\nu}\) and \(P_L^{\mu\nu}\) are the same as in thermal equilibrium while \(C^{\mu\nu}\) and \(D^{\mu\nu}\) are new tensors that depend on the anisotropy vector \(\mathbf{n}\).
The self-energy components \(\Pi_e\), \(\Pi_L\), \(\Pi_T\) and \(\Pi_d\) are given in \cite{Hauksson:2021okc} which contains further details.

As \(\Pi_d = \mathcal{O}(\xi)\) we can approximate  
\beq
D_{\ret}^A \approx \frac{1}{\left( Q^2 - \Pi_L\right) \left( Q^2 - \Pi_c\right)}
\eeq
up to order \(\mathcal{O}(\xi)\).
We furthermore ignore \(\mathcal{O}(\xi)\) corrections in the numerator as they do not describe anisotropic corrections to screening which are the non-equilibrium corrections we focus on. Then 
\begin{align}
D_{\mathrm{ret}}^{\mu\nu}(Q) &\approx \left(P_T^{\mu\nu} - C^{\mu\nu} \right) D_{\ret}^B \nonumber \\
&\qquad + \left[\left( Q^2 - \Pi_c\right) P_L^{\mu\nu} + \left(Q^2 - \Pi_L \right) C^{\mu\nu} \right] D_{\ret}^A \nonumber \\
&= \frac{P_T^{\mu\nu} - C^{\mu\nu}}{Q^2 - \Pi_e} + \frac{P_L^{\mu\nu}}{Q^2 - \Pi_L} + \frac{C^{\mu\nu}}{Q^2 - \Pi_c} 
\end{align}
The terms with the tensor \(C^{\mu\nu}\) go like
\beq
C^{\mu\nu} \left[ \frac{-1}{Q^2 - \Pi_e} + \frac{1}{Q^2 - \Pi_c}\right] = C^{\mu\nu} \frac{\Pi_c - \Pi_e}{(Q^2 - \Pi_e)(Q^2 - \Pi_c)}
\eeq
which can be ignored as \(\Pi_c - \Pi_e = \mathcal{O}(\xi)\) in the numerator.
We are thus left with 
\beq
\label{Eq:Dret_appr}
D_{\ret}^{\mu\nu}(Q) \approx \frac{P_T^{\mu\nu}}{Q^2 - \Pi_e} + \frac{P_L^{\mu\nu}}{Q^2 - \Pi_L}.
\eeq
at small anisotropy where we only include anisotropic corrections in the denominators.
The tensors \(P_L\) and \(P_T\) are the same as in equilibrium while \(\Pi_e\) and \(\Pi_L\) have anisotropic corrections. (We call \(\Pi_e = \Pi_T\) in the main text of the paper.)

\section{Details of numerical method}
\label{sec:app_num}

In this Appendix we discuss how to solve Eqs. \eqref{Eq:f11z} and \eqref{Eq:f11y}  numerically. Unlike the isotropic equation,  Eq. \eqref{Eq:f0_eq}, the anisotropic equation has a non-vanishing source term on the right hand side for all values of \(b\). One thus needs a different numerical solution method than that developed in \cite{Aurenche:2002wq, Aurenche:2002pd,Jeon:2003gi} for the equilibrium case. We note that the differential equation for the function \(f_1^{(3)}\) is 
\begin{align}
\label{Eq:f13}
\mathcal{K} \left[ f_1^{(3)}(\mathbf{b}) \right] & + \frac{ik}{2p(p+k)} \frac{8}{b^2} f_1^{(3)}(\mathbf{b})
+\mathcal{C}_0(b) f_1^{(3)}(\mathbf{b})\nonumber \\
&=  -\frac{1}{2} \mathcal{C}_1^{(b)}(b) f_0
\end{align} 
but we will not discuss this function further as it is not needed to evaluate \(A_z\) and \(A_y\) in Eqs. \eqref{Eq:Az_impact} and \eqref{Eq:Ay_impact}

For concreteness, we focus on solving Eq. \eqref{Eq:f11z}.
Defining a scaled function
\beq
G = \frac{\pi}{2} \frac{k}{p(p+k)m_{\infty}^2} \frac{f_1^{(1z)}(b)}{b},
\eeq
as well as scaled collision kernels \(\overline{\mathcal{C}}(t) = \frac{2p(p+k)}{k} \frac{1}{m_{\infty}^2} \mathcal{C}(b)\) and  variable \(t = m_{\infty} b\), this equation becomes
\begin{align}
\label{Eq:fb0}
-i &\left[ \frac{d^2 G}{d t^2} + \frac{3}{t} \frac{d G}{dt} -  G\right]   + \overline{\mathcal{C}}_0(t) G \nonumber \\
&= - \left[\overline{\mathcal{C}}_1^{(a)}(t) + \frac{1}{2}\overline{\mathcal{C}}_1^{(b)}(t)\right] \overline{f}_0(t)
\end{align}
where \(\overline{f}_0(t) = \frac{\pi}{2} \frac{k}{p(p+k)m_{\infty}^2} \mathbf{b}\cdot\mathbf{f}_0/b^2\). As shown in \cite{Aurenche:2002pd}, we can write \(\overline{f}_0 (t) = K_1(t)/t + \overline{f}_0^{\mathrm{rest}}(t)\)  where \(\overline{f}_0^{\mathrm{rest}}(t)\) is function that is finite in the limit \(t\rightarrow 0\) and which we know numerically using the methods of \cite{Aurenche:2002wq, Aurenche:2002pd}.


We need to solve Eq. \eqref{Eq:fb0}, imposing the boundary conditions that \(G(t) \rightarrow 0\) as \(t\rightarrow \infty \), as well as that \(G(t)\) is finite as
\(t\rightarrow 0\).  These boundary conditions are difficult to satisfy simultaneously for a numerical solution.
Instead we find a numerical solutions \(g(t)\) of Eq. \eqref{Eq:fb0} with \(g(t\rightarrow \infty) = 0\) and a solution of the homogeneous equation without the source term that also satisfies \(g_0(t\rightarrow \infty) = 0\). In general, both \(g(t)\) and \(g_0(t)\) blow up as \(t\rightarrow 0\). However, we know that the solution we are seeking can be written as
\beq
G(t) = g(t) + A g_0(t)
\eeq
where \(A\) is chosen so that \(G(0)\) is finite. 

We can find an explicit expression of \(A\) by using analytic solutions of Eq. \eqref{Eq:fb0} for \(t\ll 1\). In that limit the right hand side is
\begin{eqnarray}
\label{Eq:ab_def}
&&\left[\overline{\mathcal{C}}_1^{(a)}(t) + \frac{1}{2}\overline{\mathcal{C}}_1^{(b)}(t)\right] \overline{f}_0(t) \nonumber \\
&&\approx \left[\overline{\mathcal{C}}_1^{(a)}(t) + \frac{1}{2}\overline{\mathcal{C}}_1^{(b)}(t)\right] K_1(t)/t  \approx a + b 
\log t
\end{eqnarray}
where \(a\) and \(b\) are constants that depend on the momenta \(k\) and \(p\) as well as the masses \(m_D^2\) and \(m_{\infty}^2\). The differential equation becomes 
\beq
-i \left[ \frac{d^2 g}{d t^2} + \frac{3}{t} \frac{d g}{dt} -  g\right]  + \overline{\mathcal{C}}_0(t) g = a + b\log t
\eeq
which has general solution
 \beq
 \label{Eq:alpha_def}
 g(t) = w(t) + \alpha_1 w_1(t) + \alpha_2 w_2(t)
 \eeq
where 
\beq
\label{Eq:wt}
w(t) = -ia  - ib (\log t + \frac{2}{t^2})
\eeq
is an exact particular solution and \(w_2(t) = 2 J_1(it)/(it)\) and \( w_1(t) = \frac{\pi}{2}Y_1(-i t)/(-i t)  - \frac{1}{4} \left( 2 \gamma_E - 2 \log 2 + i\pi\right) w_2(t) \)
 are solutions of the homogenous equation such that \(w_1(t) = 1/t^2 + \frac{1}{2}\log t + \mathcal{O}(1)\) and \(w_2(t) = \mathcal{O}(1)\) for small \(t\).
 Similarly, the homogeneous equation can be written as 
 \beq
 \label{Eq:beta_def}
 g_0(t) = \beta_1 w_1(t) +  \beta_2 w_2(t).
 \eeq
The coefficients \(\alpha_1\), \(\alpha_2\), \(\beta_1\) and \(\beta_2\) can be found from Eqs. \eqref{Eq:alpha_def} and \eqref{Eq:beta_def} by equating the numerical solutions \(g(t)\) and \(g_0(t)\) and their derivatives with the analytic functions at some small value \(t_{\mathrm{min}} \ll 1\).

The small \(t\) behaviour of Eqs. \eqref{Eq:alpha_def} and \eqref{Eq:beta_def} shows that 
\beq
\label{Eq:Asol}
A = \frac{2ib - \alpha_1}{\beta_1}
\eeq
which makes
\beq
\label{Eq:G0_result}
G(0) = -ia - \frac{1}{2}ib + \alpha_2 + \frac{(2ib - \alpha_1)\beta_2}{\beta_1}
\eeq
finite. This is the expression that we need in order to evaluate Eq. \eqref{Eq:Az_impact}. All quantities are known for numerical solutions \(g(t)\) and \(g_0(t)\).

Eq. \eqref{Eq:G0_result} suffers from numerical cancellation errors in the terms \(\alpha_2 - \alpha_1 \beta_2 / \beta_1\). It is thus better to rewrite these terms using Eqs. \eqref{Eq:alpha_def} and \eqref{Eq:beta_def} which shows that
\beq
\label{Eq:alpha_rewriting}
\alpha_2 - \frac{\alpha_1 \beta_2}{\beta_1} = \frac{g^{\prime }g_0 - g g_0^{\prime} + g_0^{\prime }w - g_0 w^{\prime}}{g_0 w_2^{\prime} - g_0^{\prime} w_2}.
\eeq
where all quantities are evaluated at \(t_{\mathrm{min}} \ll 1\). The culprit behind cancellation errors is the term \(g^{\prime }g_0 - g g_0^{\prime}\). It can be evaluated more precisely by noting that 
\beq
\mathcal{G}_W(t) = g^{\prime } (t)g _0(t) - g(t) g_0^{\prime}(t).
\eeq
solves the equation
\beq
\label{Eq:GW_evol}
\frac{d (t^3 \mathcal{G}_W)}{dt} =  -i t^3 \left[\overline{\mathcal{C}}_1^{(a)}(t) + \frac{1}{2}\overline{\mathcal{C}}_1^{(b)}(t)\right] \overline{f}_0(t) g_0
\eeq
which can be integrated to give \(\mathcal{G}_W(t_\mathrm{min})\) and thus a reliable value of \(G(0)\).


\bibliography{refs}

\end{document}